\documentclass[prl,letterpaper,twocolumn,showpacs,floatfix,preprintnumbers,superscriptaddress]{revtex4-1}
\usepackage{algorithm,algorithmicx,algpseudocode}
\usepackage{dcolumn,amsmath,amssymb,graphicx,color,placeins}
\usepackage{algorithmicx}
\usepackage{pdfpages}

\newcommand{\gras}[1]{\boldsymbol{#1}}
\newcommand{\UNEDF}{{\textsc{unedf}}}

\newcommand{\UNEDFTWO}{{\textsc{unedf2}}}
\newcommand{\UNEDFONE}{{\textsc{unedf1}}}
\newcommand{\UNEDFONEanl}{{\textsc{unedf1}$_{\rm CPT}$}}

\begin{document}

\date{\today}

\title{Uncertainty Quantification for Nuclear Density Functional Theory and
Information Content of New Measurements}

\author{J.D. McDonnell}
\affiliation{Department of Physics and Astronomy, Francis Marion University,
Florence, South Carolina 29501, USA}
\affiliation{Physics Division, Lawrence Livermore National Laboratory,
Livermore, California 94551, USA}

\author{N. Schunck}
\affiliation{Physics Division, Lawrence Livermore National Laboratory,
Livermore, California 94551, USA}

\author{D. Higdon}
\affiliation{Los Alamos National Laboratory, Los Alamos, New Mexico 87545, USA}

\author{J. Sarich}
\affiliation{Mathematics and Computer Science Division, Argonne National
Laboratory, Argonne, Illinois 60439, USA}

\author{S.M. Wild}
\affiliation{Mathematics and Computer Science Division, Argonne National
Laboratory, Argonne, Illinois 60439, USA}

\author{W. Nazarewicz}
\affiliation{Department of Physics and Astronomy and NSCL/FRIB Laboratory,
Michigan State University, East Lansing, Michigan  48824, USA}
\affiliation{Physics Division, Oak Ridge National Laboratory, Oak Ridge, Tennessee 37831, USA}
\affiliation{Institute of Theoretical Physics, Faculty of Physics, University of Warsaw, Warsaw, Poland}

\begin{abstract}

Statistical tools of uncertainty quantification can be used to assess the
information content of measured observables  with respect to present-day
theoretical models; to estimate model errors and thereby improve predictive
capability; to extrapolate  beyond the regions reached by experiment; and to
provide meaningful input to applications and planned measurements.  To showcase
new opportunities offered by such tools, we make a rigorous analysis of
theoretical statistical uncertainties in nuclear density functional theory using Bayesian
inference methods. By considering the recent  mass measurements from the
Canadian Penning Trap at Argonne National Laboratory, we demonstrate how the
Bayesian analysis and a direct least-squares optimization, combined with
high-performance
computing, can be used to assess the information content of the new
data with respect to a model based on the Skyrme  energy density functional
approach. Employing the posterior probability distribution computed with a
Gaussian process emulator, we apply
the Bayesian framework to propagate theoretical statistical uncertainties in
predictions of nuclear masses, two-neutron dripline, and fission barriers.
Overall, we find that the new mass measurements do  not impose a constraint that
is strong enough to lead to significant changes in the model parameters. The
example discussed in this study sets the stage for quantifying and maximizing
the impact of new measurements with respect to current modeling and guiding future experimental efforts, thus
enhancing the experiment-theory cycle in the scientific method.
\end{abstract}


\maketitle

%
%
\textit{Introduction} -- Our understanding of heavy, complex nuclei lies at the
heart of many basic science questions, such as chemical evolution, neutron star
structure, synthesis of superheavy elements, mechanism of nuclear fission, or
search for the new Standard Model \cite{Decadal2012,*RISAC};
this knowledge is also crucial for  societal applications
\cite{aliberti2006,*abdel-khalik2008,*salvatores2008,*palmiotti2013}. In all
those cases, reliable theoretical estimates of nuclear masses, low-lying
excitations, electromagnetic strength, and nuclear reaction rates form
essential inputs when direct experimental information is not available.

For tackling complex nuclei theoretically, nuclear density functional theory
(DFT) is the
microscopic tool of choice \cite{bender2003}. In recent years, largely because
of algorithmic developments and high-performance computing
\cite{bogner2013}, DFT has taken great
strides as a predictive theory that describes the properties of nuclei across the nuclear
landscape \cite{Goriely09,erler2012,(Afa13)}. No consensus exists, however, on
the form of the nuclear effective interaction or
energy density functional (EDF),  resulting in large systematic
uncertainties.
Moreover, nuclear EDFs are characterized by coupling constants that must be
adjusted to experiment \cite{bender2003,(Klu09),dobaczewski2014,schunck2014}.
The systematic calculation of
uncertainties related to the determination of model parameters, as well as the
propagation of these uncertainties in model prediction, has thus become a
necessity
\cite{Rei10,gao2013,(Kor13),goriely2014,agbemava2014,dobaczewski2014} (see also
\cite{ISNET}).
Furthermore, as we enter the era of experiments with exotic nuclei at
extremes of isospin, theory will play an increasingly important role in
identifying scientific priorities  of planned experimental campaigns.
Conversely, as experiments extend current knowledge by providing information
about the  uncharted regions of the nuclear landscape, new
methodologies become critical for evaluating the
impact of these measurements on theory.

From the viewpoint of statistics, determining the parameters of a model
given a set of experimental data measurements is an inverse
problem \cite{tarantola2005}. Bayesian inference methods \cite{Box1992} are one of the
most popular and powerful statistical approaches to inverse problems, with
diverse applications in physics \cite{Dose03,RevModPhys.83.943}
(for recent nuclear physics applications, see, e.g., Refs.~\cite{Kawano06,Ireland08,Steiner10,herman2011,szpak2011,Pratt14,Graczyk14}).
In the Bayesian setting, model parameters are treated as random variables,
and their uncertainty is characterized by their joint probability distribution.
Various techniques, often based on Monte Carlo simulations,
have been developed to reconstruct this probability distribution
from model prediction of experimental data.

\textit{Objectives} -- In this work, we present the advanced application of Bayesian
inference to global nuclear properties  using nuclear DFT. In particular, we use
the Bayesian framework to quantify and propagate
DFT statistical model uncertainties and to assess the information content of
new data with respect to model developments. To this end,
we study the impact of the recently reported mass measurements from  the Canadian Penning Trap (CPT) mass spectrometer
at Argonne National Laboratory~\cite{Savard2006,vanschelt2012,vanschelt2013}
 on the Bayesian posterior probability distribution as well as the
direct determination of EDF parameters.
The CPT dataset is unique in that it probes neutron-rich nuclei around $^{132}$Sn;
hence, it can help improve our knowledge of isovector  EDF properties and
reduce extrapolation uncertainties into the region of the astrophysical
r-process.
From the resulting posterior distribution, we assess model
 uncertainties on observables, including the position of the two-neutron
dripline and fission barrier heights of actinide nuclei.

%
%
\textit{Method} -- Our theoretical framework is nuclear
density functional theory with Skyrme EDFs. Pairing is
modeled with a density-dependent pairing force and treated at the
Hartree-Fock-Bogoliubov (HFB) level by using an approximate particle number
projection with the Lipkin-Nogami method. We choose the {\UNEDFONE}
parameterization of the Skyrme functional as our reference model
\cite{kortelainen2012}. This EDF is characterized by twelve parameters that
were optimized on a set of binding energies for spherical and deformed
nuclei, charge radii, odd-even mass differences, and excitation energies of
selected fission isomers (see
Refs.~\cite{kortelainen2010,kortelainen2012,kortelainen2014} for details of the
model and the {\UNEDF} EDF family).

The quality of the functional is measured by a composite $\chi^{2}$ function,
\begin{equation}
\chi^{2}(\gras{x})
=
\frac{1}{n_{d} - n_{x}} \sum_{t=1}^{n_{T}} \sum_{j=1}^{n_{t}}
\left( \frac{y_{tj}(\gras{x}) - d_{tj}}{\sigma_{t}} \right)^{2},
\label{eq:chi2}
\end{equation}
where $\gras{x}$ denotes the set of model parameters, $n_{x} = 12$ is the number of
model parameters, $n_{T}$ the number of different data types used in the fit
($n_{T} = 4$ in our case), $n_{t}$ is the number of data points used for each
data type, $n_{d} = \sum_{t} n_{t}$ is the total number of data points, and
$d_{tj}$ and $y_{tj}(\gras{x})$ are the experimental value and corresponding
model prediction, respectively, for the $j$th data point of type $t$.
For the {\UNEDFONE} functional, where $n_{d} = 115$,
computing the $\chi^{2}$ requires about 5 minutes of CPU time
with over 800 cores in a multithreaded implementation of
the DFT solver HFBTHO \cite{stoitsov2013}.
Monte Carlo simulations used to construct
the posterior distribution may typically involve tens of thousands of such
$\chi^{2}$ evaluations; even with current supercomputers, this cost is too high.
We thus replace the DFT model $y_{tj}(\gras{x})$ with a Gaussian process (GP)
response surface, allowing Monte Carlo--based Bayesian computation.

The GP response surface is estimated within the encompassing Bayesian
formulation \cite{higdon2008cmc} by using an ensemble of DFT runs
for each of the $n_d$ experimental nuclei used in (\ref{eq:chi2}). The ensemble
is defined by a $200 \times n_x$ matrix of input settings distributed according
to a space-filling Latin hypercube sample \cite{Ye:Li:Sudj:algo:2000} over an
$n_x$-dimensional hyperrectangle centered on the {\UNEDFONE} values.
For each parameter, widths are determined
according to the standard deviations reported in Ref.~\cite{kortelainen2012},
which were obtained through a covariance analysis that assumed a linear
approximation. The GP is controlled by a scaling parameter, as well as
correlation parameters regulating the smoothness of the response surface
in each of the $n_{x}$ parameter directions.

The full posterior density includes a likelihood term for the experimental
data based on Eq.~(\ref{eq:chi2}) and the ensemble of training runs for the
GP, the uniform prior for the model parameters $\gras{x}$, and priors
for the parameters that control the GP-based response surface; see
Ref.~\cite{higdon2014} for a detailed description
of the posterior density.  We construct dependent samples from this
distribution using Markov chain Monte Carlo as detailed
in \cite{higdon2008cmc}, from which summaries such as
90\% probability intervals and posterior means can be constructed.

%
%
\textit{Results} -- Through Bayesian model calibration, we first obtained the
posterior probability distribution for the {\UNEDFONE} parameter set, which
provides a sense of how the set of fit observables of {\UNEDFONE}  constrains
the parameters.
\begin{figure}[bt]
\includegraphics[width=1\columnwidth]{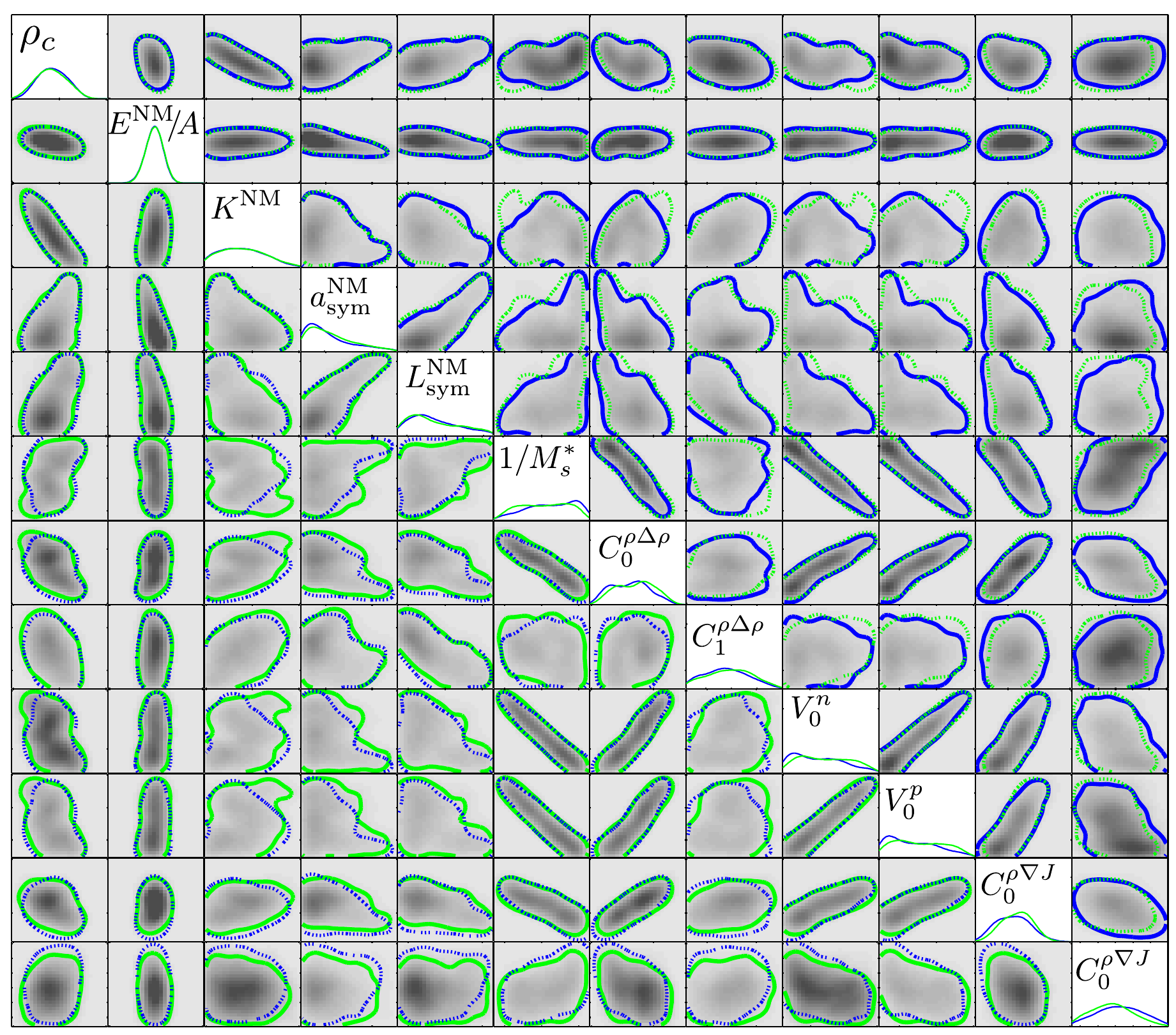}
\caption{(Color online) Univariate and bivariate marginal estimates of the posterior
distribution for the 12-dimensional DFT parameter vector
of the
{\UNEDFONE} parameterization. The blue lines enclose an estimated 95\% region
for the
posterior distribution found when only the original {\UNEDFONE} data are
accounted for; the green-outlined regions represent the same region for the
posterior distribution found when the new CPT  mass measurements are
included. The ranges of parameter variations are
 $0.155 \leq \rho_{\rm c} \leq 0.165$ (fm$^{-3}$);
 $ -16.0 \leq E^\text{NM}/A \leq -15.5$ (MeV);
 $200\leq K^\text{NM} \leq 240$ (MeV);
 $28 \leq a_\text{sym}^\text{NM} \leq 30$ (MeV);
  $20 \leq L_\text{sym}^\text{NM} \leq 60$ (MeV);
  $0.8\leq 1/M_{s}^{*}\leq1.2$;
 $-60\leq C_{0}^{\rho\Delta\rho} \leq-40$ (MeV\,fm$^5$);
  $-200\leq C_{1}^{\rho\Delta\rho} \leq-90$ (MeV\,fm$^5$);
  $-200\leq V_0^n\leq -150$ (MeV\,fm$^3$);
  $-220\leq V_0^p\leq -180$ (MeV\,fm$^3$);
  $-80\leq C_{0}^{\rho\nabla J} \leq-60$ (MeV\,fm$^5$);
  and
    $-80\leq C_{1}^{\rho\nabla J} \leq0$ (MeV\,fm$^5$).
}
\label{fig:unedf1Posterior-withoutArgonne}
\end{figure}
In Fig.~\ref{fig:unedf1Posterior-withoutArgonne}, we show the univariate and
bivariate
marginal estimates of the posterior distribution. The blue-outlined regions
give the $95$\% posterior probability region for the original {\UNEDFONE}
parameters. We notice that the Bayesian approach is in agreement
with estimates of uncertainties based on covariance analysis reported in
Ref.~\cite{kortelainen2012}. In particular, most distributions are centered on
the {\UNEDFONE} values, and the standard deviations extracted from the distribution
are consistent with the 95\% probability intervals.

In a second step, we used our Bayesian formulation to evaluate the information
content of the new mass measurements~\cite{Savard2006,vanschelt2012,vanschelt2013}.
To this end, we modified the $\chi^{2}$ of Eq.~(\ref{eq:chi2}) to include 17 new
masses of neutron-rich even-even nuclei measured at the CPT; the experimental
values are listed in the supplemental material \cite{supplemental}. The GP
response surface was again produced by using an augmented ensemble of
$200\,(n_d+17)$ DFT model evaluations. The green-outlined regions in
Fig.~\ref{fig:unedf1Posterior-withoutArgonne} represent the same 95\%
posterior probability regions obtained with the inclusion of the Argonne mass
measurements. With the
exception of a few ill-constrained parameters (e.g., nuclear incompressibility
and
isovector surface coupling constant), the shift in the posterior is small for
each parameter. This suggests a weak impact of the additional data on our model.

\begin{table}[!ht]
\caption{Root-mean-square deviations for each of the types of data included in
the {\UNEDF} optimization. Masses and energies are in MeV, radii in fm.}
\begin{ruledtabular}
\begin{tabular}{lcc}
Class          &  {\UNEDFONE} & {\UNEDFONEanl} \tabularnewline
\hline
masses (def) &   0.721            &   0.578               \tabularnewline
masses (sph)  &   1.461            &   1.545               \tabularnewline
radii       &   0.022            &   0.022               \tabularnewline
odd-even staggering (n)        &   0.023            &   0.024               \tabularnewline
odd-even staggering (p)        &   0.079            &   0.081               \tabularnewline
fission isomer energies          &   0.190            &   0.316               \tabularnewline
masses (CPT)     &   1.064            &   0.479             \tabularnewline
\end{tabular}
\end{ruledtabular}
\label{tab:RMS_UNEDF1-ANL}
\end{table}

For comparison, we performed a direct reoptimization,
independent of the GP response surface,
of the {\UNEDFONE}
functional that includes the new CPT masses \cite{WSS14}. We refer to the
reoptimized EDF parameter set as {\UNEDFONEanl}; see supplemental
material for parameter values \cite{supplemental}. The two parameterizations
are similar. The largest relative difference, weighted
by the standard deviations reported in Ref.~\cite{kortelainen2012}, is
0.6$\sigma$ for the isovector surface coupling constant $C_{1}^{\rho\Delta\rho}$ and surface symmetry
energy. These quantities have been difficult to constrain with the
dataset used in the {\UNEDF} protocol. Of interest, then, is the fact
that the {\UNEDFONEanl} value of $C_{1}^{\rho\Delta\rho}$
is close to that of {\UNEDFTWO}, which was also optimized to effective
single-particle energies \cite{kortelainen2014}, known to be sensitive probes
of surface properties.
Since the new dataset including CPT
masses is more skewed toward neutron-rich nuclei, it may supply additional
information about the shell structure above
doubly magic $^{132}$Sn through a better determination of isovector coupling
constants.

Table \ref{tab:RMS_UNEDF1-ANL} displays the root-mean-square deviation
between calculated and measured values for each type of data included in the
optimization. We note that the inclusion of the CPT mass measurements shifts
the optimization  priority, so that the new masses and deformed masses are
reproduced more accurately, while predictions for fission isomers and spherical
masses deteriorate slightly. The results in Table~\ref{tab:RMS_UNEDF1-ANL}
are indicative of a small,  additional
constraint on the isovector coupling
constants in {\UNEDFONEanl}.

\begin{figure}[htb]
\includegraphics[width=1\columnwidth]{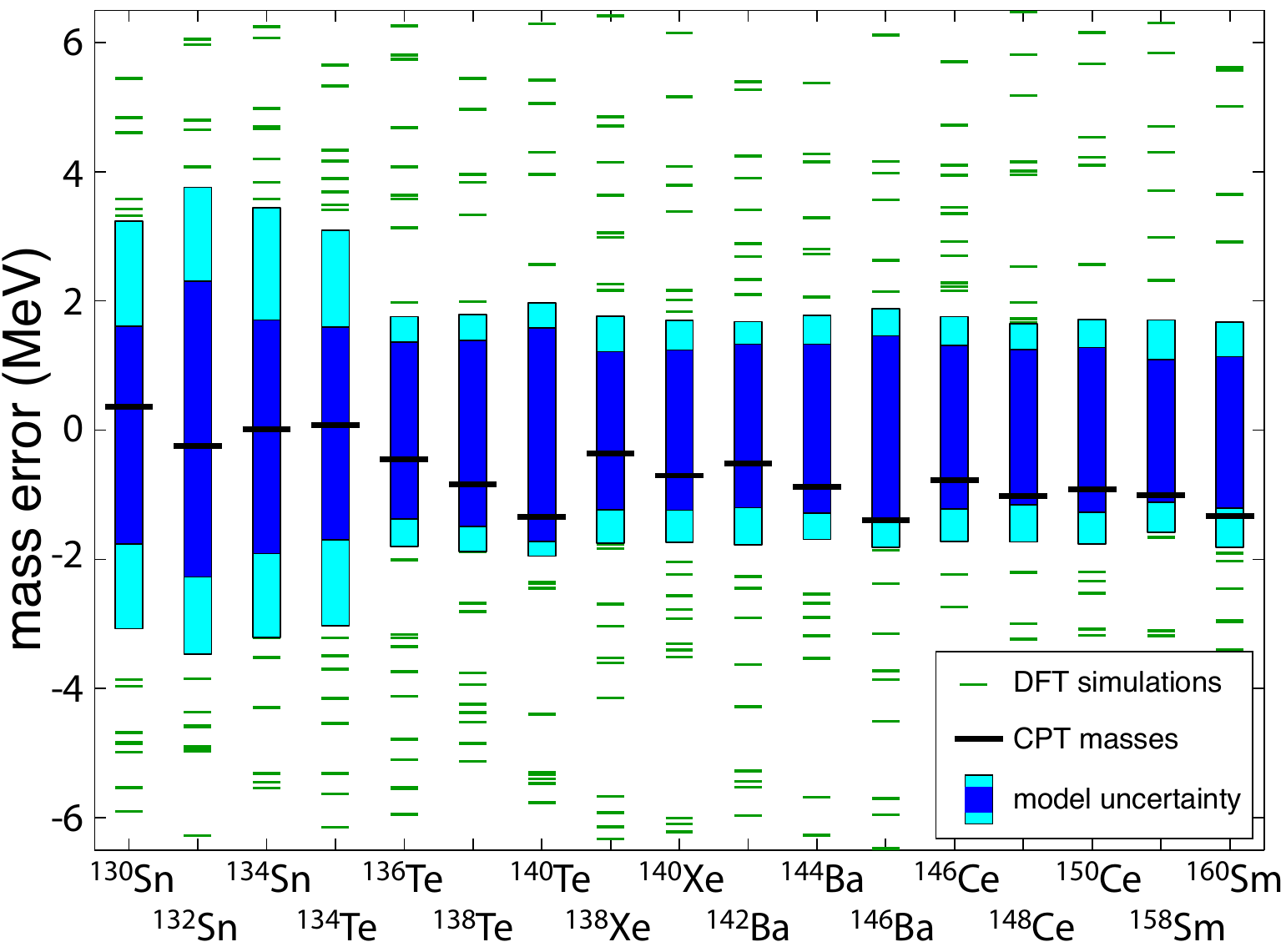}
\caption{(Color online) Estimated theoretical error bars for the masses of the
even-even nuclei measured in
Refs.~\cite{Savard2006,vanschelt2012,vanschelt2013}, using the posterior for
{\UNEDFONE}.
Dark blue bands represent the 90\% confidence bands obtained from the
posterior; larger, light blue bands also account for model error;
black bars show  mass residuals.}
\label{fig:ArgonneErrorBars}
\end{figure}
Equipped with the posterior distribution for the EDF parameters, we now turn to
the propagation of statistical uncertainties for model predictions. We take the posterior
distribution for the EDF parameters obtained by conditioning only on the
{\UNEDFONE} measurements and propagate the distribution through the augmented
GP-based emulator, producing prediction intervals for the new CPT mass
measurements.  These estimates are genuine {\em holdout} predictions since the
new  mass data were not used in determining the posterior distribution.
Figure~\ref{fig:ArgonneErrorBars} shows 90\% prediction intervals
(centered on the mean mass value of {\UNEDFONE}) for the new CPT masses. The
dark blue band is the 90\% interval for the uncertainty in the EDF model
parameters; the light blue band also includes uncertainty due to model error.
The model error uncertainty was estimated from the difference between the
posterior mean estimate and the actual mass measurements in the
{\UNEDFONE} dataset.  Separate estimates were made for spherical and
deformed nuclei.  These estimated model-error standard deviations were assumed
to be appropriate for the new CPT mass measurements, producing this additional
uncertainty.
We observe that the experimentally measured values (black bars in
Fig.~\ref{fig:ArgonneErrorBars}) are generally within the 90\%
prediction interval.
The estimated uncertainty for the calculated masses is approximately
$\pm 2$ MeV, and slightly larger for the four spherical nuclei (the first four
nuclei in the figure).
This uncertainty is relatively large and in excellent agreement with the
r.m.s.\ deviation for masses of even-even nuclei across the entire nuclear
landscape, which is 1.9\,MeV for {\UNEDFONE}.

\begin{figure}[htb]
\includegraphics[width=1\columnwidth]{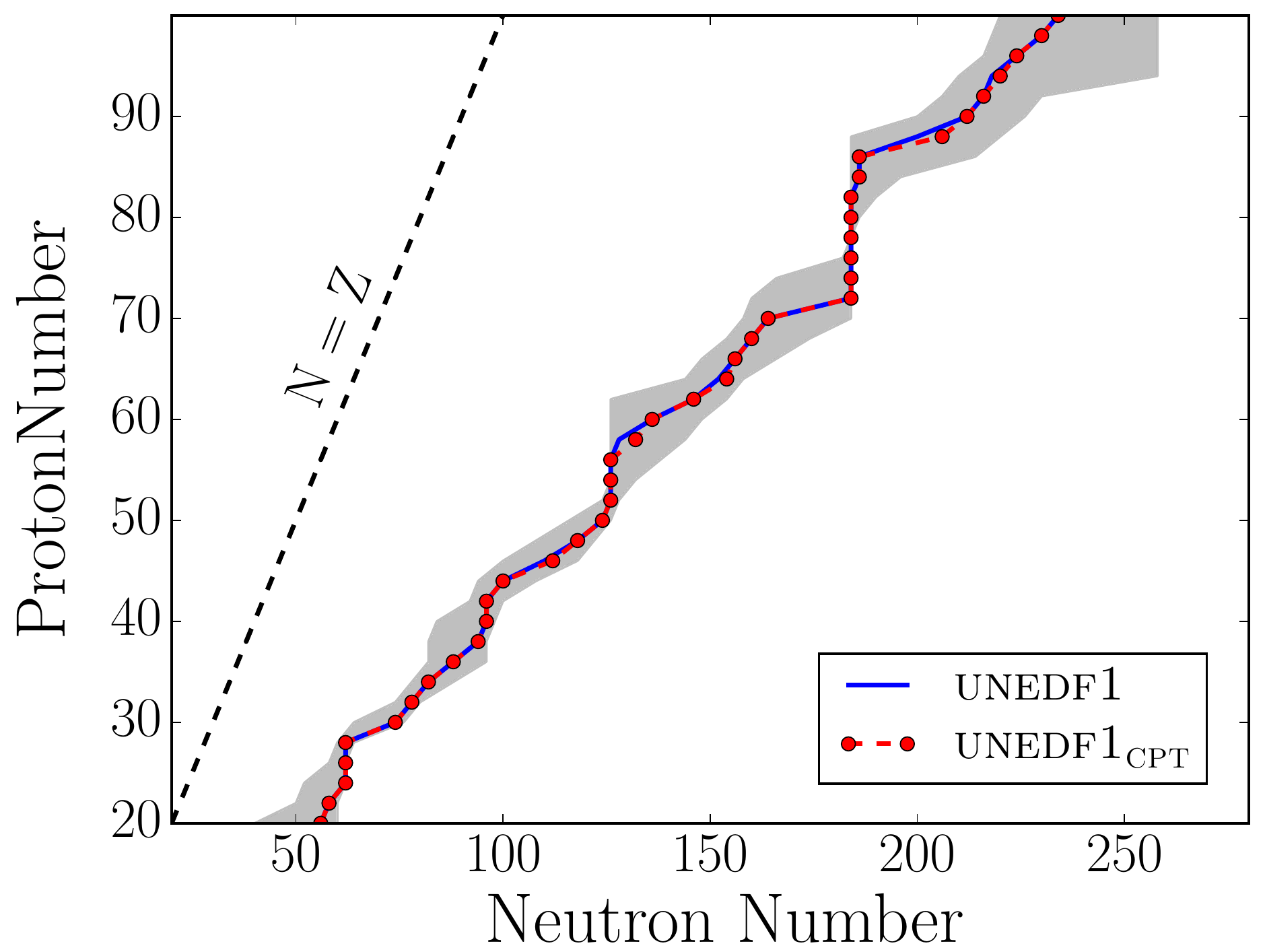}
\caption{(Color online) Comparison between the two-neutron dripline predictions made with
{\UNEDFONE} (solid line) and those made with {\UNEDFONEanl} (dashed line). The 90\% probability spread
about the {\UNEDFONE} predictions is  shown in grey.}
\label{fig:DriplineComparison}

\end{figure}

We now evaluate how the calculated model uncertainties impact predictions for
important physical observables. We first look at the position of the
two-neutron dripline, which is especially important for our understanding of
nucleosynthesis in the r-process \cite{grawe2007}. For a given element
characterized by its proton number $Z$, the two-neutron dripline is defined
as the point where the two-neutron separation energy becomes negative. We have performed an ensemble of calculations of
nuclear binding energies for all even-even neutron-rich elements with $20 \leq Z \leq 100$
over the Latin hypercube sample design of EDF parameter inputs, allowing yet another
GP-based emulator to be constructed for these binding energies.

Once the emulator is constructed, we propagate the posterior distribution of
the
model parameters (conditioning on either the {\UNEDFONE} or {\UNEDFONEanl}
datasets), producing uncertainty in the estimated dripline.  With this
Monte Carlo sample, we can estimate the posterior mode and
90\% interval for the dripline for each value of  $Z$.
We explored the axial quadrupole potential energy surface of each
nucleus to allow for deformed solutions. The results are presented in
Fig.~\ref{fig:DriplineComparison}. We observe that the inclusion of 17 new
masses of neutron-rich nuclei in the optimization protocol did not impact the
position of the dripline, since results with {\UNEDFONE} and {\UNEDFONEanl}
are practically indistinguishable. The predicted dripline is consistent with the
results of large-scale DFT surveys \cite{erler2012,(Afa13)}. Apart from the few
closed-shell, waiting-point nuclei, the uncertainty on the
position of the dripline is on the order of 15 to 20 nucleons. This is
comparable to statistical and systematic uncertainties obtained by comparing
predictions made with different Skyrme functionals \cite{erler2012}.

\begin{figure}[htb]
\includegraphics[width=1\columnwidth]{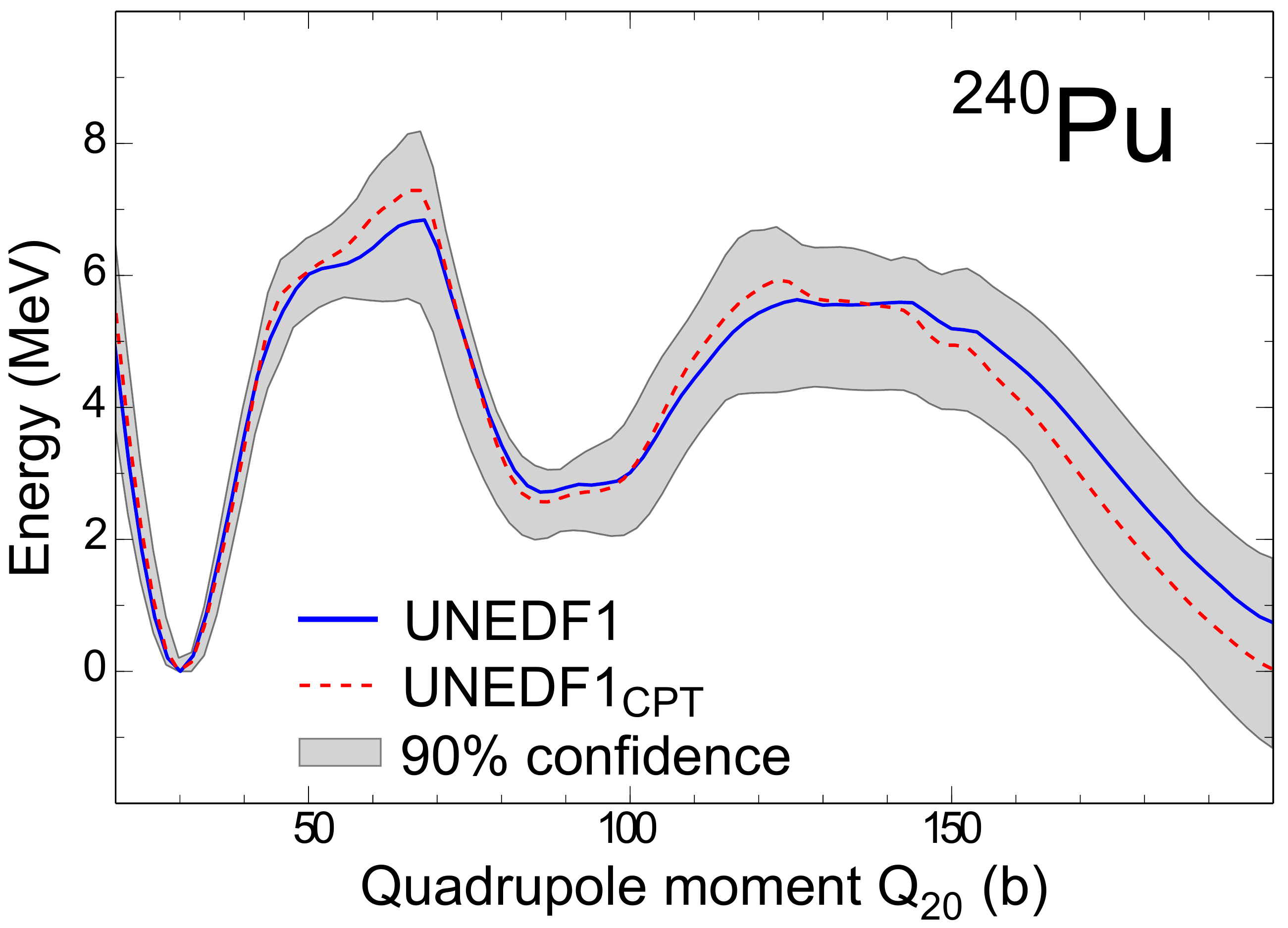}
\caption{(Color online)  Comparison between the fission barrier predictions for
$^{240}$Pu made with {\UNEDFONE} (solid line) and those made with
{\UNEDFONEanl} (dashed line), together with the 90\% confidence interval (shaded
grey area). The potential
energy surface was obtained by following the lowest-energy static fission pathway in a
four-dimensional collective space of axial and triaxial quadrupole,
axial octupole, and axial hexadecapole mass moments.
}
\label{fig:Pu240FissionComparison}
\end{figure}

Another important application area  of nuclear DFT is fission theory.
In Fig.~\ref{fig:Pu240FissionComparison}, we show the potential energy
curve of $^{240}$Pu. This nucleus is representative of the actinide region
and is often used as a theoretical benchmark.   Again, the results of {\UNEDFONE}
and {\UNEDFONEanl} are close. The large theoretical uncertainty in
the predicted static fission barrier is worth noting; similar results were
obtained in Ref.~\cite{Erlerfis} in the context of fission properties for
r-process nuclei.
Since a 1\,MeV shift in the fission barrier translates into
many  orders of magnitude difference in the spontaneous
fission half life, such results highlight the urgent need for better
constraining the deformation properties of current EDFs.

%
%
\textit{Conclusions} -- We have presented a comprehensive application of
Bayesian
inference techniques to the calculation and propagation of theoretical
statistical uncertainties in nuclear density functional theory.
By using the recent, unique dataset of mass measurements from the CPT at
Argonne
National Laboratory, we showcase how the statistical tools of uncertainty
quantification and high-performance computing can be used to assess the
information content of new data with respect to current  models. Such
analyses will become increasingly relevant for enhancing the feedback
 in the  ``observation-theory-prediction-experiment"- cycle of the scientific
method at the eve of next-generation radioactive ion beam facilities and
exascale computing.

In the particular case studied in this work, we found that the impact of
the new neutron-rich nuclei mass data on our DFT model is minor.
The coupling constants  of the earlier functional {\UNEDFONE} and of
the new functional {\UNEDFONEanl}, informed by the new data,  are fairly close;
hence, their
predictions for the two-neutron dripline and fission barrier in $^{240}$Pu
are practically identical.
Although the major theoretical statistical uncertainty in developments of
the nuclear EDF comes from the poorly constrained isovector terms and the
new data on neutron-rich nuclei are generally expected to reduce this
uncertainty, the  lack of a significant constraint
from the new masses  suggests that both the amount of new neutron-rich
isotope data and the range of neutron excess probed, are not
sufficiently large  to impact our model appreciably.
Moreover, because of their poor precision with respect to the existing data
(see Table~\ref{tab:RMS_UNEDF1-ANL}), even the current, best-calibrated EDFs
are not sensitive and flexible enough to fully take advantage of the  new experimental information.

By propagating theoretical errors, we found large model uncertainties
in the predictions of the two-neutron dripline and the fission barrier in
$^{240}$Pu. In this respect, we concur with the conclusions of
Ref.~\cite{vanschelt2013} that
existing mass models are  insufficient for accurate r-process simulations.
Clearly, accurate  measurements for nuclei with even larger neutron excess,
closer to  the r-process path, are still needed in order to better inform
theory.

We note that the uncertainties discussed in this work are estimated
statistically, reflecting parameter uncertainty and model misfit. The misfit
error is most likely due to our lack of knowledge of the form of the nuclear
EDF itself, and additional measurements will never reduce this source of uncertainty.
Adding physics that is
missing in the current implementations of nuclear DFT is a major challenge for
the field. A distinct and complementary challenge is to develop tools that deliver uncertainty
quantification for theoretical studies as well as for the assessment
of new experimental data. The present work represents a step in this direction.

%
%
\begin{acknowledgments}
We are grateful to Guy Savard and Robert Janssens for helpful discussions
and access to CPT masses.
This material is
based upon work supported by the U.S.\ Department of Energy, Office of
Science, Office of Nuclear Physics under award numbers
DE-AC52-07NA27344 (Lawrence Livermore National
Laboratory),  DE-AC02-06CH11357 (Argonne
National Laboratory), and DE-SC0008511
(NUCLEI SciDAC Collaboration), and by the NNSA's Stewardship Science
Academic Alliances Program under award no.\ DE-NA0001820.
Computational resources were provided through an INCITE award
``Computational Nuclear Structure'' by the National Center for Computational
Sciences and National Institute for Computational Sciences at Oak
Ridge National Laboratory, through an award by the Livermore Computing Resource
Center at Lawrence Livermore National Laboratory, and through an award by the
Laboratory Computing Resource Center at Argonne National Laboratory.
\end{acknowledgments}

\bibliographystyle{apsrev4-1}
\bibliography{uq_nuclear_dft.bib}

\onecolumngrid
\newpage

\includepdf[openright, fitpaper=true, pages=-]{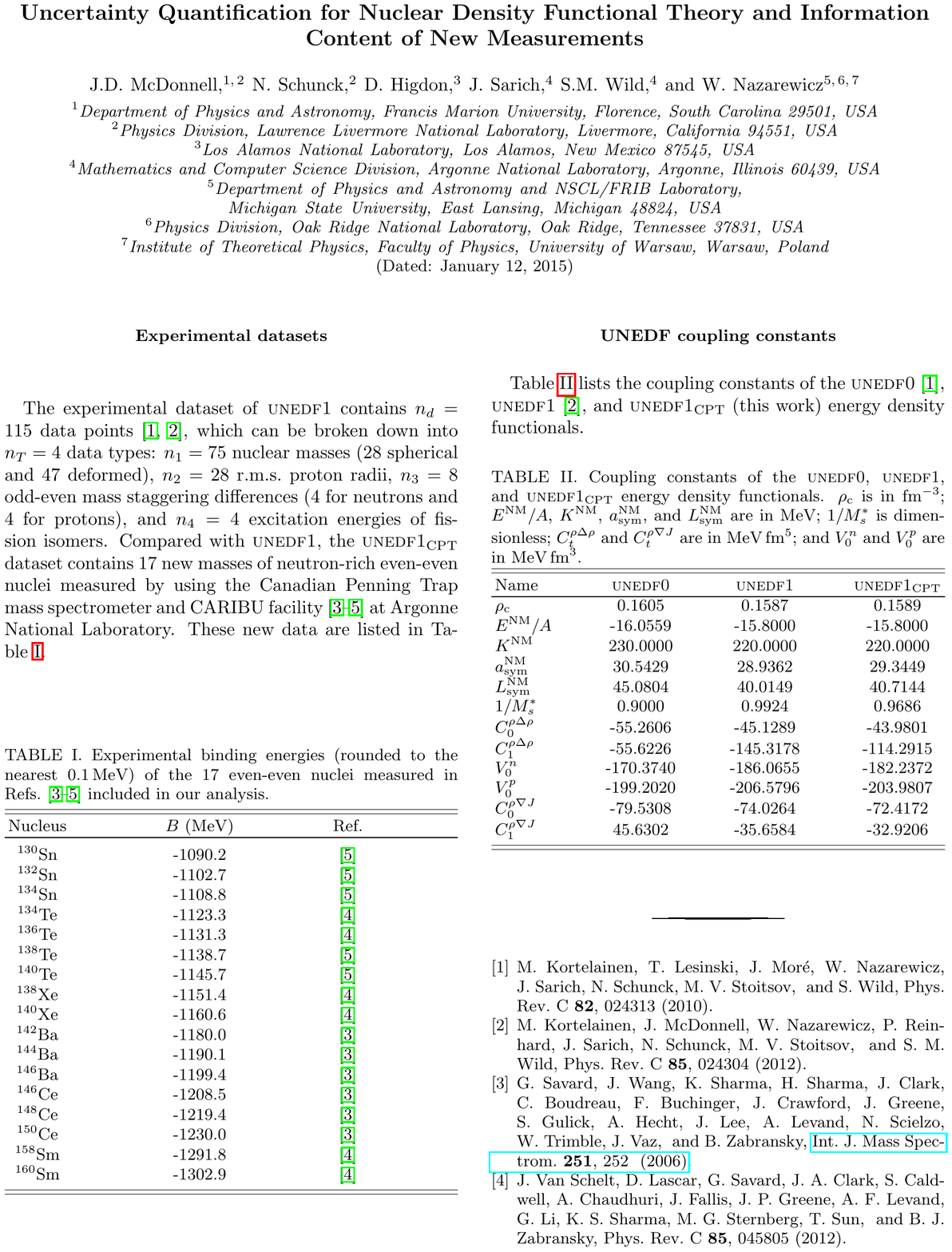}

\end{document}